\documentclass[apj]{emulateapj}

\usepackage[]{natbib}
\usepackage{graphicx}
\bibpunct{(}{)}{,}{a}{}{,}

\begin{document}

\def\bullet{{1E0657$-$56}}
\def\arcsecf{\!\!^{\prime\prime}}
\def\arcminf{\!\!^{\prime}}
\def\diff{\mathrm{d}}
\def\ngx{N_{\mathrm{x}}}
\def\ngy{N_{\mathrm{y}}}
\def\eck#1{\left\lbrack #1 \right\rbrack}
\def\eckk#1{\bigl[ #1 \bigr]}
\def\rund#1{\left( #1 \right)}
\def\abs#1{\left\vert #1 \right\vert}
\def\wave#1{\left\lbrace #1 \right\rbrace}
\def\ave#1{\left\langle #1 \right\rangle}
\def\kms{{\rm \:km\:s}^{-1}}
\def\dds{D_{\mathrm{ds}}}
\def\dd{D_{\mathrm{d}}}
\def\ds{D_{\mathrm{s}}}


\title{Strong and weak lensing united III: Measuring the mass
distribution of the merging galaxy cluster 1E0657$-$56
\altaffilmark{*}} \altaffiltext{*}{Based on observations made with the
NASA/ESA Hubble Space Telescope, obtained at the Space Telescope
Science Institute, which is operated by the Association of
Universities for Research in Astronomy, Inc., under NASA contract NAS
5-26555. These observations are associated with program \# 10200.
This work is also based in part on observations made with the Spitzer
Space Telescope and with the 6.5 meter Magellan Telescopes located at
Las Campanas Observatory, Chile. Spitzer is operated by the Jet
Propulsion Laboratory, California Institute of Technology under a
contract with NASA.}  \shorttitle{The mass distribution of \bullet}
\author{Maru\v{s}a Brada\v{c}\altaffilmark{1,2},
Douglas \ Clowe\altaffilmark{3},
Anthony \ H. Gonzalez\altaffilmark{4},
Phil \ Marshall\altaffilmark{1},
William \ Forman\altaffilmark{5},
Christine\ Jones\altaffilmark{5},
Maxim\ Markevitch\altaffilmark{5},
Scott \ Randall \altaffilmark{5}, 
Tim \ Schrabback\altaffilmark{2}, and
Dennis\ Zaritsky\altaffilmark{3}}
\shortauthors{Brada\v{c} et al.}
\defcitealias{bradac04a}{Paper I}
\defcitealias{bradac04b}{Paper II}
\altaffiltext{1}{Kavli Institute for Particle Astrophysics and Cosmology,
P.O. Box 20450, MS29, Stanford, CA 94309, USA}
\altaffiltext{2}{Argelander-Institut f\"{u}r Astronomie, Auf dem H\"{u}gel 71,
D-53121 Bonn, Germany}
\altaffiltext{3}{Steward Observatory, University of Arizona, 933 N Cherry Ave., Tucson,
AZ 85721, USA}
\altaffiltext{4}{Department of Astronomy, University of Florida, 211 Bryant Space Science Center, Gainesville, FL 32611, USA}
\altaffiltext{5}{Harvard-Smithsonian Center for Astrophysics, 60 Garden Street, Cambridge, MA 02138, USA}
\email{marusa@slac.stanford.edu}


\begin{abstract}
The galaxy cluster 1E0657-56 ($z = 0.296$) is remarkably
well-suited for addressing outstanding issues in both galaxy evolution
and fundamental physics. We present a reconstruction of
the mass distribution from both strong and weak
gravitational lensing data. Multi-color, high-resolution HST ACS
images allow detection of many more arc candidates than were
previously known, especially around the subcluster. Using the known
redshift of one of the multiply imaged systems, we determine the
remaining source redshifts using the predictive power of the strong lens
model. Combining this information with shape measurements of
``weakly'' lensed sources, we derive a high-resolution, absolutely-calibrated
mass map, using no assumptions regarding the physical
properties of the underlying cluster potential. This map provides the best
available quantification of the total mass of the central part of the cluster.
We also confirm the result from \citet{clowe04,clowe06b}
that the total mass does {\it not} trace the baryonic
mass.
\end{abstract}


\section{Introduction \label{sec:intro}}
 
The cluster of galaxies {\bullet} is one of the hottest, most X-ray
luminous clusters known. Since its discovery by \citet{tucker95}, it
has been the subject of intense and ongoing research
\citep{mehlert01,markevitch02,barrena02,clowe04,markevitch04,gomez04}. In
particular, Chandra observations by \citet{markevitch02} revealed the
cluster to be a supersonic merger in the plane of the sky with a
textbook example of a bow shock, making this cluster a unique case in
which to study hydrodynamical properties of interacting systems. The
optical images show that the cluster has two distinct components, and
the X-ray analysis reveals that the lower mass sub-cluster has
recently exited the core of the main cluster with a relative velocity
of $4500^{+1100}_{-800} \kms$. Although this relative velocity appears
to be unusually large, an analysis of cosmological simulations
demonstrates that it is well within the predicted range of the
currently favored cosmological model \citep{hayashi06}.

Due to its unique geometry and physical state, this cluster is the
best known system in which to test the dark matter hypothesis
\citep{clowe04}. The observed offsets between the weak gravitational
lensing mass peaks and the X-ray gas component give the most direct
evidence for the presence of dark matter yet available. Using the same
observations, \citet{markevitch04} placed upper limits on the dark
matter self-interaction cross section.

The goal of this work is to obtain a high-resolution, absolutely
calibrated mass map with no assumptions on the physical properties of
the underlying cluster potential. For this purpose we use HST ACS
data, incorporating the gravitational lensing information from both
multiple image systems (strong lensing) and from distortions of
background sources (weak lensing). The superb spatial resolution
delivered by ACS both increases the number density of background
sources that can be used for weak lensing and reveals new strong lensing
candidates, especially around the subcluster. The joint strong and
weak lensing analysis thus benefits greatly from the ACS data,
enabling us to increase the spatial resolution and signal strength of
the mass map in the core of the cluster.

Encouraged by the success of the
combined strong and weak lensing reconstruction method developed in
\citet{bradac04a} (hereafter \citetalias{bradac04a}) and the results from applying the method
to cluster RX~J1347.5$-$1145 \citep{bradac04b}, we proceed to apply this method to
{\bullet}. Because the strong lensing data are richer in this case
(RX~J1347.5$-$1145 has not been observed with either the ACS or WFPC2
cameras prior to this study), we improve the method as described in the text.

We perform the reconstruction in the following sequence. Using ACS
(multi-color where available) HST images, we identify the
multiply imaged systems. Having a spectroscopic redshift for one
lensed system (from \citealp{mehlert01}), we then use the predictive
power of simple strong lens modeling to estimate the redshifts of the
other systems. Using positions and redshifts of the strongly lensed images and
the shape measurements of the ``weakly'' lensing sources, we perform a
combined strong and weak lensing mass reconstruction, thereby
significantly improving the constraints on the mass and positions of
the main cluster and the colliding subcluster of {\bullet}.
In terms of the structure of this paper, in section~\ref{sec:data} we describe
the optical images used in this analysis, the basic image processing and the
extraction of the strong~(\S\ref{sec:datasl}) and
weak~(\S\ref{sec:datawl}) gravitational lensing data. 
We then infer the mass distribution of cluster {\bullet}
from these data in section~\ref{sec:swunited_bullet}, following a
demonstration of our methodology on suitable simulated data in
section~\ref{sec:swunited}. We discuss the possible sources of error in
section~\ref{sec:systematics} and summarize our conclusions in
section~\ref{sec:conclusions}.

  
\section{Observations and data reduction process}
\label{sec:data}

ACS/WFC imaging of the cluster {\bullet} was carried out in Cycle 13
(proposal 10200, PI Jones) on 2004 October 21 in two
pointings with one and three different filters respectively. The two
pointings are centered on the main cluster and the subcluster
with a small overlap between them. The subcluster  was
observed in three different filters (three orbits with F814W, one with
F606W, and one with F435W), while the main cluster was observed only
in the F606W filter (one orbit).

The demands placed by the lensing analysis require special care when
reducing the images. We use the {\tt Multidrizzle}
\citep{multidrizzle} routine to align the images. To register the
images with the astrometric accuracy needed for lensing analysis, we
determine the offsets among the images by extracting high $S/N$
objects in the individual, distortion corrected exposures. We use {\tt
Sextractor} \citep{sextractor} and the IRAF routine {\tt geomap} to
identify the objects and calculate the residual shifts and rotation of
individual exposures, which were then fed back into {\tt
Multidrizzle}. We use ``square'' as the final drizzling kernel and an
output pixel scale of $0.03\mbox{ arcsec}$; this is smaller than the
original pixel scale of the ACS CCD to reduce the impact of resampling
on the shape measurements.


\subsection{Strong lensing image identification \label{sec:datasl}}

Owing to its complex structure, the strong lensing analysis of the
{\bullet} cluster is complicated. Among the multiply-imaged systems, only the giant arc on
the NW side of the main cluster's cD has a measured redshift 
\citep[$z=3.24$][]{mehlert01}.
Corresponding images for multiply-imaged sources were identified by matching both
morphologies and surface brightnesses in each of the available
ACS bands (F435W, F606W, and F814W) and ground-based filters, 
BVR data from Magellan and I-band from VLT
\citep{clowe04}.

\begin{figure*}[ht]
\begin{center}
\includegraphics[width=1.0\textwidth]{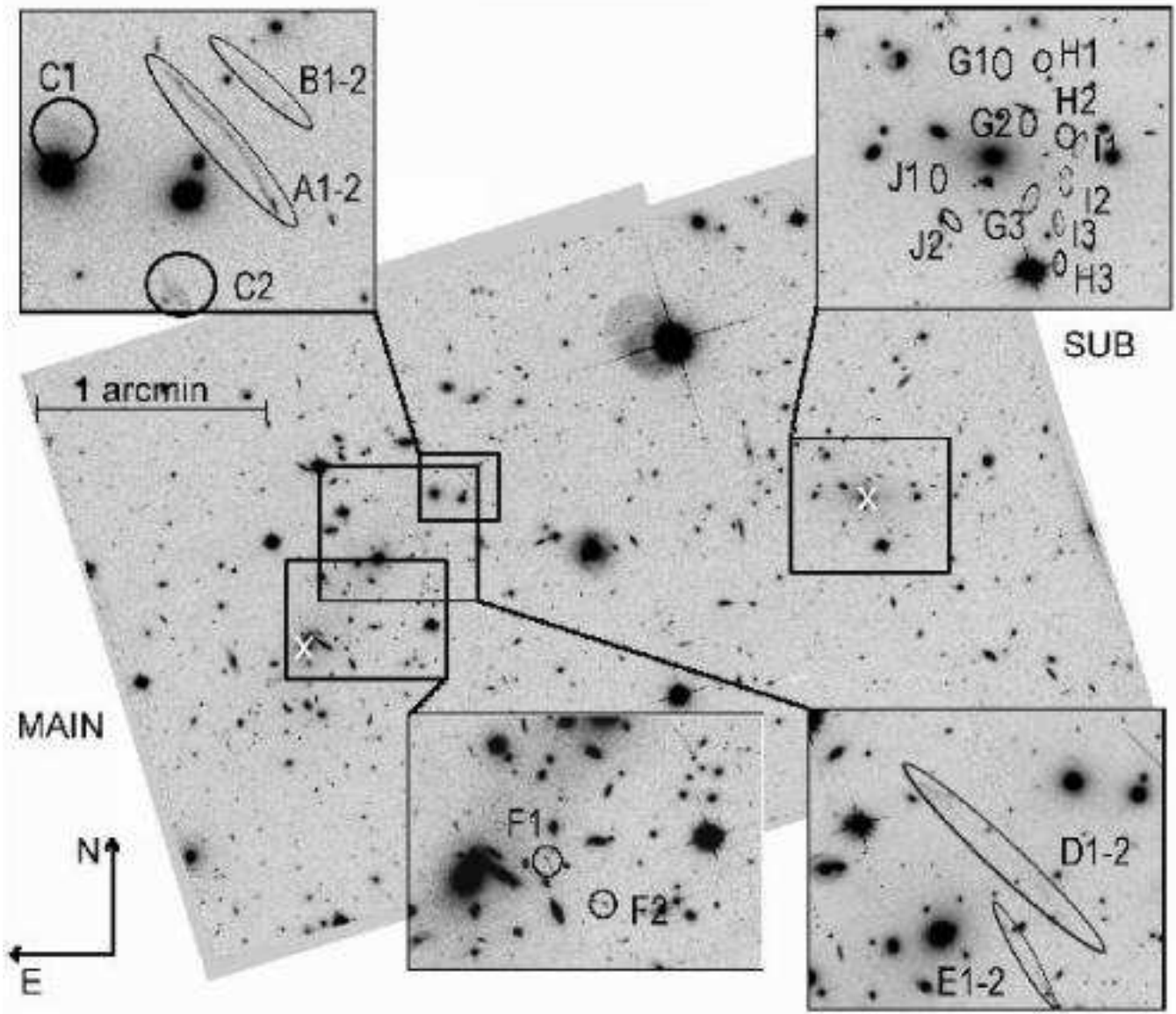}
\end{center}
\caption{The two F606W pointings of the main and sub component region of \protect \bullet. Multiply imaged systems are marked and labeled (see also \protect{Table~\ref{tab:arcs}}). White crosses denote the positions of the southern cD of the main and BCG of the subcluster.}
\label{fig:arcslabel}
\end{figure*}

We confirm (based on photometry and morphology)
that the $6$ systems (labeled A-F in Table~\ref{tab:arcs}) previously
identified by \citet{mehlert01} are indeed multiply
imaged. We identify four additional
systems (G-J) in the subcluster region, where
none were previously known. All of these are identified in
Fig.~\ref{fig:arcslabel}, and image positions (corresponding to the
peak surface brightness) are presented in Table~\ref{tab:arcs}.
Unfortunately, for the greater part of the main cluster, only
a single band ACS image is available, which makes identification of additional
candidates significantly more ambiguous.

Using these identifications, we perform a parametrized strong
lensing reconstruction. At this stage we are not
generating a detailed strong lensing model, rather we use this
parametrized model only to predict the redshifts of the systems where
spectroscopic redshifts are not available (all systems but A) and as
the initial model guess for the subsequent strong and weak lensing
reconstruction. For this reconstruction, we use only the image positions as
constraints.  The parametrized model consists of two non-singular
isothermal ellipses (NIE) \citep{ke98}, for which the scaled surface mass
density $\kappa$ is given by
\begin{equation}
  \kappa(\vec {\theta}) = \frac{b}{2\sqrt{\frac{1+\abs{\epsilon_{\rm
            g}}}{1-\abs{\epsilon_{\rm g}}} \rund{r_{\rm c}^2 +
        (\theta_1)^2} + (\theta_2)^2}} \; .
\label{eq:nis}
\end{equation}
$\vec{\theta}$ is defined counterclockwise from due west and $b$ is related
to the line-of-sight velocity dispersion $\sigma$ through $b \propto
\sigma^2$. The components are centered on the southern cD galaxy of
the main cluster and the brightest cluster galaxy (BCG) of the sub
cluster respectively (denoted with white crosses in
Fig.~\ref{fig:arcslabel}; it is not straightforward to identify the
BCG of the main cluster, therefore we refer to the two main galaxies
as the southern and northern cD). We allow the scaling $b$, core radii
$r_{{\rm c}}$, ellipticities $\abs{\epsilon_{\rm g}}$, and position angles $\phi_{{\rm g}}$ of these to vary.\footnote{We define
all ellipticities $\epsilon$ throughout this paper as in
\citet{bartelmann00} with $\abs{\epsilon}=(1-f)/(1+f)$, $f$ being the
axes ratio.}
Following the prescription of \citet{kneib96}, we also include the 20
brightest cluster members from the F606W-band (selected using color
information from Magellan data) in the mass model.  The galaxies are
modeled as non-singular isothermal spheres with a line-of-sight
velocity dispersion $\sigma_{\rm memb}$ and core radius $r_{\rm
c,memb}$ following
\begin{equation}
\sigma_{\rm  memb}  \propto L^{1/4} \; , r_{\rm c,memb} \propto
L^{1/2} \; , 
   \label{eq:23}
\end{equation}
The proportionality constants are allowed to vary as well and below
we quote the values of $\sigma_{\rm memb}$ and $r_{\rm c,memb}$ for a
fiducial galaxy with a F606W-band magnitude $m_{\rm F606W}=18$. The best fit
model for this system has values of $\{\sigma, r_{\rm c},
\abs{\epsilon_{\rm g}}, \phi_{\rm g}\} = \{1300 \kms,
0.3^{\prime}, 0.2, 45^{\circ}\}$ for the main cluster, $\{\sigma,
r_{\rm c}, \abs{\epsilon_{\rm g}}, \phi_{\rm g}\} =  \{1000\kms,
0.3^{\prime},0.2,280^{\circ}\}$ for the subcluster, and $\{\sigma_{\rm
memb, 18}, r_{\rm c,memb,18}\} = \{250\kms, 0.3^{\prime} \}$ for the
cluster members. We also allow the redshifts of systems B-J to
vary. The resulting best fit redshifts are given in
Table~\ref{tab:arcs}. To evaluate the angular diameter
distances throughout the paper we assume the $\Lambda$CDM cosmology
with $\Omega_{\rm m} = 0.3$, $\Omega_{\Lambda} = 0.7$, and Hubble
constant $H_0 = 70 {\rm \ km \: s^{-1}\:Mpc^{-1}}$.

For  objects close in projection to the main cluster, the redshift estimates are
fairly reliable because the well-measured arc A is located there (we
estimate an error of $\Delta z = 0.2(1+z)$). However, the redshift
estimates for those
close to the subcomponent are less well-determined, and are
somewhat degenerate with the adopted mass of the subcluster.  We postpone
an extension of  the methodology described in the following sections that incorporates
simultaneous solution for the unknown source redshifts to future work,
but here note that: 1) the parametrized model allows the generation of
a set of source redshifts that, while model-dependent, are
self-consistent; 2) the predicted redshifts are consistent with
the colors of the objects.  We adopt the redshifts estimated here for
the combined strong and weak lensing reconstruction, and where
possible estimate the systematic error introduced as a result of this simplification.

\begin{deluxetable}{rrrr} 
\tablecolumns{4} 
\tablewidth{0pc} 
\tablecaption{The properties of the multiply-imaged systems used in this work.}
\tablehead{ \colhead{} &  \colhead{Ra} & \colhead{Dec} & \colhead{$z_{\rm pred}$}}
\startdata
& 104.63316 & -55.941395 & \\
\raisebox{0.5ex}{A} & 104.62988 & -55.943798 & \raisebox{0.5ex}{3.24\footnote{The redshift
  of  arc A was determined spectroscopically by \citet{mehlert01}. Other redshifts are determined using the best-fit strong lensing model.}} \\
\cline{1-4}
& 104.62954 & -55.941844 & \\
\raisebox{0.5ex}{B} & 104.63042 & -55.941474 & \raisebox{0.5ex}{4.8} \\
\cline{1-4}
& 104.63775 & -55.941851 & \\
\raisebox{0.5ex}{C} & 104.63338 & -55.945324 &  \raisebox{0.5ex}{2.1} \\
\cline{1-4}
& 104.64709 & -55.943575 & \\
\raisebox{0.5ex}{D} & 104.63528 & -55.951836 &  \raisebox{0.5ex}{1.4} \\
\cline{1-4}
& 104.64008  &  -55.950620 & \\
\raisebox{0.5ex}{E} & 104.64232  &   -55.948784&  \raisebox{0.5ex}{1.0} \\
\cline{1-4}
&104.65155 & -55.956671& \\
\raisebox{0.5ex}{F} & 104.64778 & -55.958284 &  \raisebox{0.5ex}{0.8} \\
\cline{1-4}
& 104.56568  & -55.939832 & \\ 
G & 104.56402& -55.942113 &1.3\\
& 104.56417 & -55.944131  & \\
\cline{1-4}
 & 104.56293 & -55.939764 & \\
H & 104.56133 & -55.942430 & 1.9\\
 & 104.56189 & -55.947724 & \\
\cline{1-4}
&104.56186& -55.946114  & \\
I &104.56052 &-55.942930 & 2.1\\
&104.56141 &-55.944264 & \\
\cline{1-4}
& 104.56909 & -55.946016  & \\
\raisebox{0.5ex}{J} & 104.57025 & -55.944050 & \raisebox{0.5ex}{1.7} 
\enddata 
\label{tab:arcs}
\end{deluxetable} 


\subsection{Weak lensing catalogs \label{sec:datawl}}

In the weak lensing analysis, we use the F435W, F606W, and F814W
exposures for the subcluster, and F606W wherever multicolor ACS data
are not available. We will only outline the main steps used for
this analysis, full details on how the weak lensing catalogs were
generated will be given in a future paper dealing with the weak
lensing mass measurements of {\bullet} at large radius \citep[in
prep.]{clowe06c}.

We correct galaxy images for the PSF anisotropy and PSF smearing
closely following the technique described in \citet{clowe06}. The
procedure is based on the KSB algorithm \citep{ksb95}, in particular
we use the modified {\tt IMCAT} implementation. The KSB method is
formally valid only in the weak lensing regime (i.e. at radii much
larger than where multiply imaged systems form and where the
distortions are small). However recent simulations from the STEP
project \citep{heymans06} show that this approach is also valid in the
non-weak lensing regime (i.e. with shear values of $\abs{\gamma} \sim
0.1-0.3$) to the accuracy needed for a single cluster
reconstruction. Lastly, in the vicinity of the highly elongated arcs
we are dominated by the strong lensing signal, thereby minimizing the
effects of signal dilution due to imperfect PSF correction.

We select stars from the half-light-radius vs. magnitude diagram and
fit a fifth order polynomial to their measured ellipticities as a
function of their position on the drizzled image. After the correction
the rms of the stellar ellipticity components $\epsilon_1$ and
$\epsilon_2$ (as defined in \citet{bartelmann00}) changes from $0.017$
to $0.015$ and from $0.004$ to $0.003$, respectively, and the mean is
shifted to $0$ in both cases. The shear is then measured
independently in all available filters and a weighted average is
calculated from those to arrive at our final, PSF corrected
ellipticity estimates. We use the $\sim 1^{\prime}$ overlap between
the two F606W pointings to test for systematics in the PSF
correction. The differences in the shears derived from the two
exposures is consistent given the noise level.

The ACS camera has a PSF that varies both spatially and
temporally. While the approach described above accounts for the
spatial dependence, it does not account for time variations. We
investigate the PSF time dependence by measuring stellar shapes in
individual exposures (four exposures were taken within each orbit). We
conclude that the PSF is {\it not} stable (i.e. the global pattern of
the PSF changes) even within a single orbit.  However our detected
signal is much stronger than the typical residuals due to imperfect
PSF removal caused by the temporal variation; therefore we use the
combined images to measure shapes. In addition, the PSF shape of an
individual star changes as a function of its radius (see
e.g. \citealp{heymans05,jee06}); we therefore match the radius used to
measure the stellar correction factors with that for the
galaxies. Using simulated weak lensing data we identified a
constant bias between the measured and estimated shear values; this bias was calculated for each image set separately and was introduced 
as a multiplicative shear calibration factor when producing final catalogs.

The color cuts applied to each catalog to remove cluster and
foreground galaxies were color-color cuts estimated using
photometric-redshift templates of galaxies at redshifts $z \lesssim
2$.  We used templates of all but extreme starburst galaxies, under
the assumption that there are not likely to be many faint,
low-redshift starburst dominated galaxies, particularly in a cluster.
As such, we expect that unless there is an unknown population of dwarf
galaxies in the cluster with colors much different from known
populations, we will have removed all but a handful of outliers
(extreme starbursts or dust obscured). We only use objects which
were detected in more than a single band; in particular we reject all
the objects that are detected only in the F606W pointing of the main
cluster and were too faint for detection in the Magellan images.

 Unfortunately the observations at hand do not allow us to determine
the photometric redshifts of the sources used for weak and strong
lensing. We would need at least NIR data to reliably determine the
redshifts, as most of the sources are located at $z\sim1$. Therefore
following \citet{clowe06}, we estimate the redshifts of the background
sources by applying the color and magnitude cuts that remove the
likely foreground population to the HDF-S photometric redshift catalog
of \citet{fontana99}. For the remaining sources, we average the ratios
$\dd\dds/\ds$ (where $\dd$, $\ds$, and $\dds$ are the angular diameter
distances to the cluster, source and between the cluster and the
source respectively) to generate a very crude estimate of the
individual redshifts, but good average redshifts per population
(provided the redshift distribution of the HDF-S is representative of
that of our background sources). As a result all of the galaxies
in a certain magnitude bin will be given the same redshift.  The main
reason of this approach, instead of calculating a mean for the whole
catalog is the fact that the images have different depths in different
parts of the reconstructed area, and this method allows us to account
for the sudden jumps in the mean background galaxy redshift across the
image boundaries, which would otherwise introduce spurious features in
the reconstructions.

This method is only approximate, therefore we test the reconstruction
method using mock catalogs with only limited knowledge of the redshift
of the source population in Sect.~\ref{sec:swunited}. The weak and
strong lensing reconstruction performs well under these conditions. In
addition, any significant dilution of the signal will be compensated
by the strong lensing signal (if more than one strong lensing system
is used). The cluster is also at a moderate redshift, therefore
changing the sources from redshifts of $z=1$ to $z=1.5$ would change
the lensing strength by $\sim 14\%$. In conclusion we expect the
contributions due to errors in the assumed distribution of
$\dd\dds/\ds$ to be smaller than that expected from the galaxy
ellipticity shot noise.

Finally, we fit the non-singular isothermal sphere NIS model to the
 the averaged tangential ellipticities $\ave{\epsilon_{\rm t}}$.  Both
 components generate large tangential shear signal.  We center
 circular bins on the southern cD of the main cluster and the BCG of
 the subcluster.  The resulting NIS fit for the main cluster results
 in $\sigma_{\rm NIS} = 1300 \mbox{ kms}^{-1}$, $x_{c,\rm NIS}
 =0.\arcminf4$, and $\sigma_{\rm NIS} = 1000 \mbox{ kms}^{-1}$,
 $x_{c,\rm NIS} =0.\arcminf6$ for the subcluster. This is an
 oversimplified model because the cluster is morphologically disrupted
 and we do not correct for the contamination of the main cluster
 profile in the sub cluster region (and vice versa).

\begin{figure*}[ht]
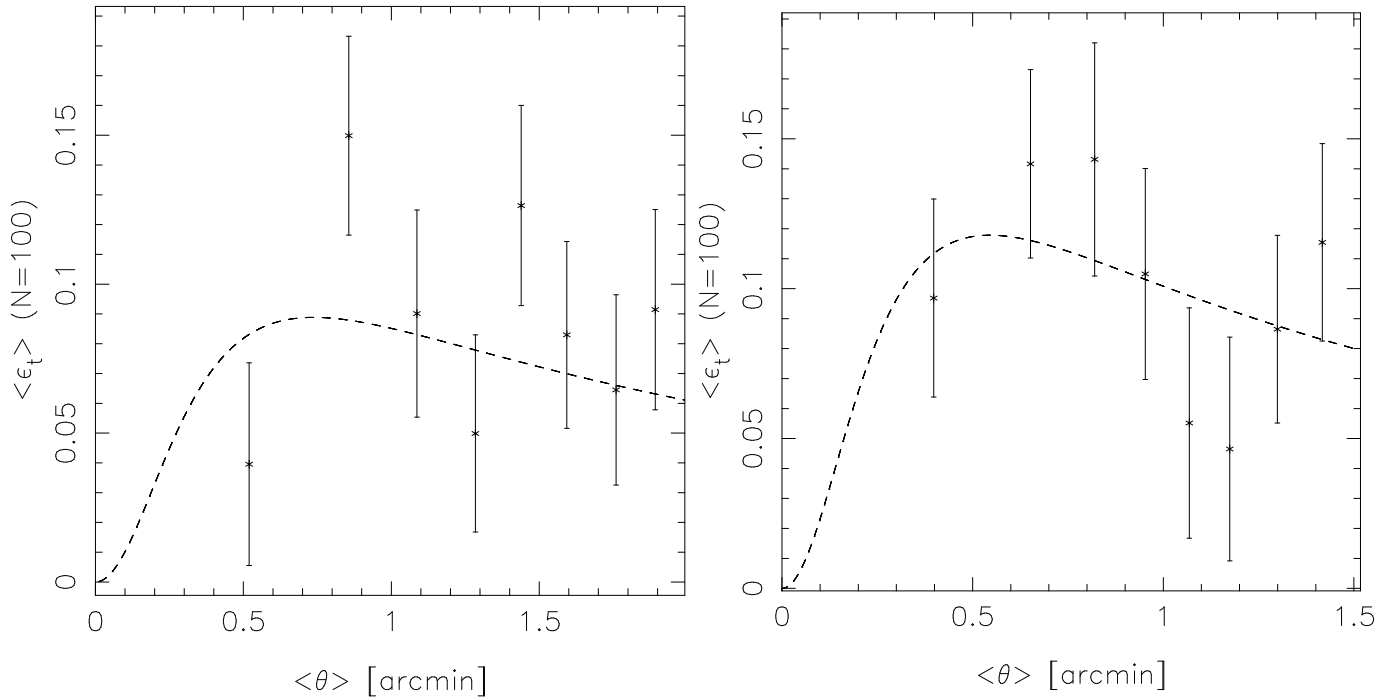

\begin{minipage}{0.5\textwidth}
\begin{center}
\includegraphics[width=1.0\textwidth]{f2a.eps}
\end{center}
\end{minipage}
\begin{minipage}{0.5\textwidth}
\begin{center}
\includegraphics[width=1.0\textwidth]{f2b.eps}
\end{center}
\end{minipage}
\caption{Average tangential ellipticity $\ave{\epsilon_{\rm t}}$ vs. projected
  radius $\ave{\theta}$ in radial bins centered on the southern cD of
  the main cluster ({\it left pannel}), and the BCG of the subcluster ({\it right pannel})
  containing 100 galaxies per bin. The errors are obtained by
  randomizing the phases of the measured ellipticities, while
  preserving their absolute values. 100 realizations are
  performed. The dashed line shows the best-fit non-singular
  isothermal sphere (NIS) profile to the binned data (see text). It is
  plotted here for the average source redshift of weak lensing
  sources $\ave{z} = 1.0$.}
\label{fig:wlprof}
\end{figure*}

\section{Strong and weak lensing reconstruction from high resolution (ACS) data}
\label{sec:swunited}

Our strong and weak lensing reconstruction is performed using the
method described in \citetalias{bradac04a}. The main idea behind the
method is to describe the cluster mass distribution using a set of
generic model parameters, which we choose to be the projected
gravitational potential $\psi_k$ measured on a regular square grid.
The total $\chi^2$ is defined as
\begin{equation}
  \label{eq:13a}
  \chi^2(\psi_k) = \chi_{\epsilon}^2(\psi_k) +
  \chi_{\rm M}^2(\psi_k) + \eta R(\psi_k) \; .
\end{equation}
where $\chi_{\epsilon}^2(\psi_k)$ is the contribution from weak
lensing, $\chi_{\rm M}^2(\psi_k)$ from strong lensing and $R(\psi_k)$
is the regularization function with regularization parameter
$\eta$. We closely follow the methodology presented in
\citetalias{bradac04a}, with the few differences described below. \footnote{In figure 2 in \citetalias{bradac04a} the factor of $\frac{1}{2}\frac{1}{6}$ in finite differencing formula for $\kappa$ is wrong. All the calculations in all three papers do however use the correct factor of $\frac{1}{6}$.}

The most important difference is that we extend the formalism
to more than a single set of multiple images. This may appear
straightforward because we only need to sum up the contributions to
$\chi_{\rm M}^2$ from each system. Given $N_{\rm S}$ multiply imaged systems having $N_i$ images each, we define $\chi_{\rm M}^2$ as
\begin{equation}
  \label{eq:17b}
\chi_{\rm M}^2 = \sum_{i=1}^{N_{\rm S}} \sum_{m=1}^{N_i} \vec b_{m,i}^{\rm\: T}
\:\mathcal{S}^{-1} \:\vec b_{m,i}\; ,
\end{equation}
where $\mathcal{S}$ is the covariance matrix and $\vec b_{m,i} = \vec
\theta_{m,i} - Z(z_{{\rm s},i}) \vec \alpha(\vec \theta_{m,i}) - \vec
\beta_{{\rm s},i}$.  $\vec \beta_{{\rm s},i}$ is the average position
of the $i$-th source (calculated from the image positions and given
the deflection angle $\vec \alpha$), and $Z(z_{{\rm s},i})$ is the
so-called ``cosmological weight'' function as defined in
\citet{bartelmann00}.

In \citetalias{bradac04a} we assumed that the covariance matrix is
diagonal and independent of the lens model parameters. As discussed in
e.g. \citet{kochanek04_sf}, this approach is not optimal and can lead
to solutions that are biased toward models that predict high
magnification at the image positions. With more sources at differing
redshifts, this problem becomes more apparent because the method 
likely converges toward a solution with steep gradients in the
surface mass density close to image positions (allowing for the ``fine
tuning'' of the reconstruction). To overcome such unphysical
solutions, we use a better approximation for the error containing the magnification at the image
position $\mu(\vec \theta_{m,i})$ and obtain the covariance matrix
$\mathcal{S} = \abs{\mu(\vec \theta_{m,i})}^{-2} {\rm
diag}(\sigma_{\rm s,1}^2,\sigma_{\rm s,2}^2)$. Such a procedure is
approximately correct, provided we are close enough to the true model
(see \citealp{kochanek04_sf}). The covariance matrix is still
diagonal; however we argue that assuming it to be such in the source
plane is in fact a better approximation than assuming the covariance
matrix to be diagonal in the image plane, as sources are on average
more circular than their lensed images. In addition, in practice
multiple image constraints are satisfied nearly perfectly and exact
values of errors on image positions are of lesser importance.

We adapt the method to accommodate data regions of arbitrary size and
shape, but the potential is reconstructed only in grid cells
containing weak and/or strong lensing data. We tile the observed field
with a regular grid and determine which cells contain either weak
lensing galaxies or multiple images. This can result in some cells
being sparsely populated (holes in the data, edge effects), but the
regularization ensures that potential is properly reconstructed. Once
the data-grid is determined, we ensure that each of the data grid
points is surrounded by a $4 \times 4$ grid of points that are
included in the reconstruction to calculate the scaled surface mass
density $\kappa$, shear $\gamma$, and the deflection angle $\vec
\alpha$, using finite differencing, as described in
\citetalias{bradac04a}.

To find the $\chi^2$-function minimum, we search for a solution
of the following system of equations $\frac{\partial
\chi^2(\psi_k)}{\partial \psi_k} = 0$. This is in general a non-linear
set of equations, which we solve in an iterative manner. We again
linearize this system as described in \citetalias{bradac04a},
calculating all of the non-linear terms (for the strong lensing term in
particular these are now the ones containing $\vec \beta_{{\rm s},i}$
and $\mu(\vec \theta_{m,i})$) using the information from the previous
iteration. We regularize the solution using a ``moving prior'',
gradually updating our knowledge of the model by increasing its
complexity and computing its likelihood relative to the previous
(simpler) model. We begin with an initial model ($\kappa^{(0)}$, $\gamma_1^{(0)}$, $\gamma_2^{(0)}$) computed on a relatively coarse grid, and then
gradually increase the number of grid points, comparing the resulting
$\kappa$ map with that from the previous iteration, linearly
interpolated onto the finer grid. In addition to $\kappa$ we now also
use the components of the shear $\gamma_1$ and $\gamma_2$ in the
regularizing scheme, to avoid the numerical effects described in
\citetalias{bradac04a} even more efficiently. The regularization
constant was set after experimentation with simulated data. When using
many multiply imaged systems, the final reconstruction is even less
sensitive to these choices than when only a single system is used (as
in \citetalias{bradac04a}); we discuss this finding in more detail in Sect.~\ref{sec:systematics}.

With many multiply imaged systems, we must choose a sufficiently
fine grid to be able to calculate the difference in the deflection
angle between images of different systems (if images from more than
one system lie within the same grid cell, it is difficult to find a
good solution). Ideally one would like to use
adaptive grids for this purpose (having a relatively coarse grid where
only weak lensing information is available and increasing the
resolution in the vicinity of the multiple images). Unfortunately it
is computationally difficult to track such a problem and is therefore
beyond the scope of this paper, but will be a subject of future work.

We tested all the improvements described above using a high-resolution
N-body simulation of a galaxy cluster by \citet{springel01}. The
catalogs are generated as in \citetalias{bradac04a}, although adapted
to the higher-resolution data that we use here.  The weak
lensing simulated data are obtained by placing $1800$ galaxies (giving a
density of $120\mbox{ arcmin}^{-2}$) randomly on a $3.8 \times 3.8
\mbox{ arcmin}^2$ field. The intrinsic ellipticities $\epsilon_{\rm s}$ are drawn from
a Gaussian distribution, with each component characterized by $\sigma
= \sigma_{\epsilon^\mathrm{s}} = 0.2$.  We then draw the redshifts of the
background sources, following \citet{brainerd96}, from a gamma
distribution $ p_\mathrm{z}(z) = {z^2} / {\rund{2\:z_0^3}}
\exp\rund{-z/z_0}$ with $z_0 = 1 / 3$, giving a mean redshift of $\langle
z \rangle = 3 z_0 = 1$, and a mode of $z_\mathrm{mode} = 2 z_0 = 2 /
3$. We reject the foreground objects, and use the background object
redshifts when predicting the lensed shapes. However, when using the
catalogs for the reconstruction, we assume {\it no} knowledge of the
individual redshifts: instead all sources are assumed to have $z=1$,
in agreement with the poor estimates available for the actual data. The multiple-image
systems are generated as described in \citetalias{bradac04a}. Four quadruply-imaged systems are
generated at source redshifts of $z_{\rm s} = \{0.8,1.0,3.0,5.0\}$. The initial model for the regularization 
is set to $\kappa^{(0)} = \gamma_1^{(0)} = \gamma_2^{(0)} = 0$.

\begin{figure*}[ht]
\begin{center}
\includegraphics[width=1.0\textwidth]{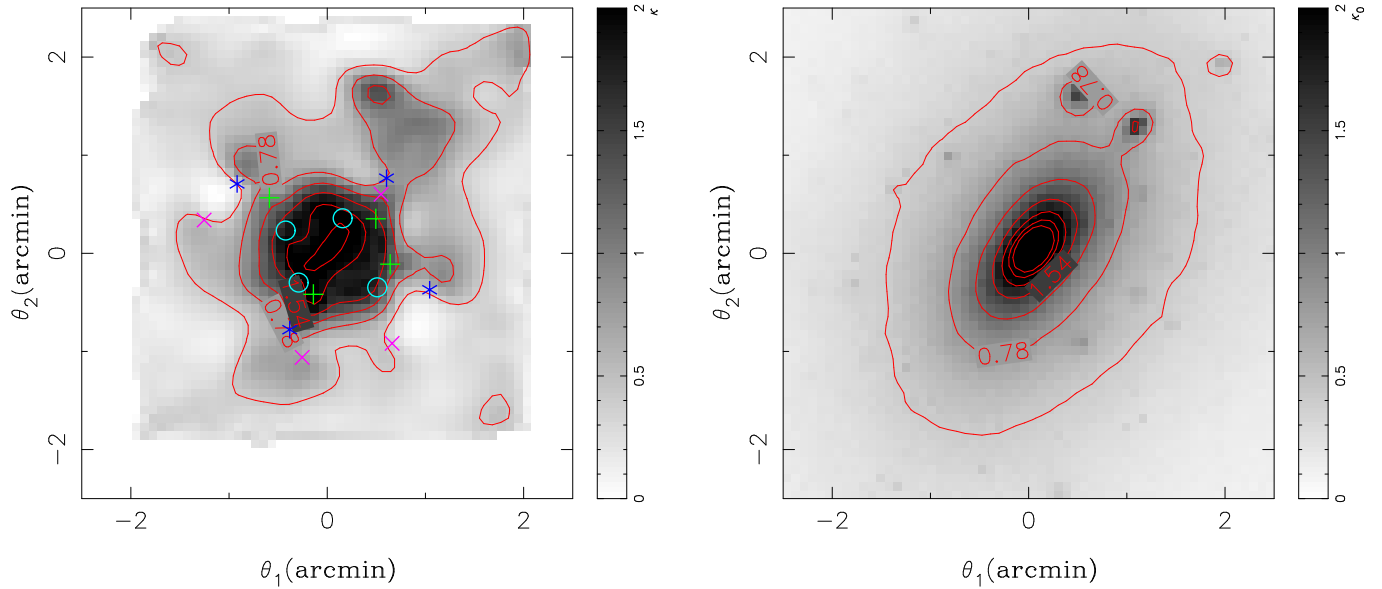}
\end{center}
\caption{The reconstructed (left) and the original (right) surface
  mass density $\kappa$ of the simulated cluster described in
  section~\ref{sec:swunited}. 
  The symbols in the left panel denote the image positions
  of the four multiple-image systems at $z_{\rm s} =
  \{0.8,1.0,3.0,5.0\}$ which we use for the reconstruction. The contour levels, which are 
  the same in both panels, are linearly spaced with
  $\Delta\kappa = 0.38$, starting at $\kappa=0.4$ for a fiducial
  source at infinite redshift, $z_{\rm s} \to \infty$.}
\label{fig:sims}
\end{figure*}

The resulting reconstruction is given in Fig.~\ref{fig:sims}. It is
very encouraging to observe how well the overall shape (the agreement
in the ellipticity and position angle of the main component, and
the overall profile) and its mass are reconstructed. These successes would not
be possible with weak lensing data alone, owing to the mass-sheet
degeneracy (i.e. the reconstruction would suffer from the degeneracy
of the form $\kappa \to \kappa' = \lambda \kappa + (1 - \lambda)$,
where $\lambda$ is an arbitrary constant, see e.g. \citealp{bradac03b}). In addition with the
ACS-quality data, we are starting to resolve the small-scale
substructure of the potential.

\section{Cluster mass reconstruction of \protect \bullet}
\label{sec:swunited_bullet}

We now apply the above mentioned reconstruction method to the strong
and weak lensing data of {\bullet}. As described in
Sect.~\ref{sec:datasl}, we use 10 distinct multiply imaged systems and
$N_{\rm gal} \sim 1900$ ``weakly'' lensed galaxies ($120\mbox{
arcmin}^{-2}$). We start with a $30 \times 30 \mbox{ pix}^2$ grid for
a $7\times 7 \mbox{ arcmin}^2$ field oriented with respect to the
negative RA coordinate (not all of the grid cells contain data and
those that do not are excluded from the reconstruction). We gradually
increase the number of grid points (for reasons described in
Sect.~\ref{sec:swunited}) in steps of 1, with the final reconstruction
performed on a $60 \times 60 \mbox{ pix}^2$ grid. In our initial
model, we use a two-component NIE model with parameters as given in
Sect.~\ref{sec:datasl} to set the regularization. The possible
dependence of the reconstruction on initial conditions is discussed in
the next section.

The resulting reconstruction is shown in Fig.~\ref{fig:swunited},
together with X-ray surface brightness contours from the 500 ks Chandra
observation. The two cluster components are clearly
detected. In addition, with superb strong lensing data (due to
exquisite ACS resolution and multi-color information for the
subcluster) we are starting to resolve multiple components within each
individual system.

From this reconstruction, we measure the enclosed (projected within a
cylinder) mass for this system. The main cluster has $M_{\rm
main}(<250\:\mbox{kpc})= (2.8 \pm 0.2) \times 10^{14} M_{\odot}$,
while the sub cluster has $M_{\rm sub}(<250\:\mbox{kpc})= (2.3 \pm
0.2) \times 10^{14} M_{\odot}$ ($1\mbox{ arcmin} = 260\mbox{ kpc}$ at
the cluster redshift). The mass errors represent combined errors
obtained by bootstrap resampling of the weak lensing catalogs,
performing a similar experiment for the strong lensing data, and
estimating systematic uncertainties from the mass differences obtained
using different initial models (see discussion in the next section).
We plot the integrated, azimuthally-averaged mass profiles $M(<R)$ for
the two components in Fig.~\ref{fig:profile}.  We fit power-law models
($M(<R) \propto R^{n_{\rm i}}$) to these profiles, and find
logarithmic slopes of $n_{\rm i,main} = 0.8$ for the main cluster and
$n_{\rm i,sub} = 1.1$ for the subcluster.  In terms of the surface mass density $\kappa(r) \propto r^{-n}$, this translates to
$n_{\rm main} = 1.2$, and $n_{\rm sub} = 0.9$. Adopting an isothermal
profile ($n = 1$), we find the line of sight velocity dispersions for
the main and the sub cluster are $\sigma_{\rm main} = (1400 \pm 100)
\kms$ and $\sigma_{\rm sub} = (1200 \pm 100) \kms$ respectively;
in good agreement with the analysis in
Sect.~\ref{sec:datawl}. However, as already mentioned in that section,
given the size of the errors, the asphericity of the system, and the
complicated geometry (the relative contributions of the main and sub
cluster are difficult to disentangle), the velocity dispersion and
slope measurements should only be used as guidelines.

Because of this cluster's violent history, we can not expect the
assumption of isothermality to hold. Therefore (and for the
reasons stated above) it is also non-trivial to compare our
velocity dispersion measurements with the ones from
\citet{barrena02}. Their estimate for the main cluster agrees with
ours, while the velocity dispersion of a subcluster is much lower in
their case. An isolated lens with the low velocity
dispersion measured by
\citet{barrena02} for the subcluster would have multiply
imaged systems only at radii $\lesssim 1.^{\arcsecf}3$, 
a factor of $\sim 8$ below that observed.
The influence of the main cluster can increase the radial range over
which multiple images exist, but
not to the extent observed here. We conclude that the true
velocity dispersion of the subcluster is likely to be smaller than that derived from
our mass model (due to the inclusion of mass from the main cluster), but
larger than that presented by \citet{barrena02}.

One of the important results of the work of \citet{clowe04,clowe06b}
was the measurement of the offset between dark matter and baryons. 
We confirm these results: our strong and weak lensing
offsets for the mass peaks (by measuring the peaks in
$\kappa$-distribution) and the corresponding errors are given in
Table~\ref{tab:offset}. To estimate the significance of this
offset we also measured the positions of the two peaks (for the main
and the subcluster) in each of the 1000 reconstructions using the
bootstrap resampled weak lensing catalogs (described
below). In none of these realizations are the mass peaks coincident with the
corresponding X-ray peaks. The 1-$\sigma$ error bars are a factor of
$10$ and $6$ smaller than the offsets for the main and subcluster
respectively. If these errors are accurate and Gaussian distributed,
this result translates into 10-$\sigma$ and 6-$\sigma$ discrepancies
between the corresponding gas and the total mass peaks. These offsets
directly demonstrate that dark matter is present and is the
dominant mass component of this cluster.

Lastly, we compare the projected gas mass (the details of
modeling will be given in \citealp{randall06}) vs. the total mass
within elliptical regions centered on the two peaks of the X-ray
emission and the southern cD of the main and BCG of the subcluster,
respectively (see Table~\ref{tab:massgas}).
The ratio of gas mass to total mass is smaller in the regions centered
on the massive galaxies rather than the X-ray peaks, indicating that the gas has
been (to some extent) stripped, and is no longer coincident with the
total mass. In addition, we detect an extension in the total mass map
to the NW of arc A. This detection can be partly attributed to the
hot X-ray gas (see Fig.~\ref{fig:swunited}).

\begin{deluxetable}{rrr} 
\tablecolumns{3} \tablewidth{0pc} 
\tablecaption{Offsets (with 1-$\sigma$ errorbars) of the mass peaks and gas peaks from the corresponding southern cD of the main and BCG of the subcluster (marked with crosses in \protect{Fig.~\ref{fig:arcslabel}}).}
\tablehead{ \colhead{} & \colhead{$\Delta x ([\mbox{arcmin}])$} &  \colhead{$\Delta y ([\mbox{arcmin}])$}} 
\startdata 
Main component & $0.26 \pm 0.08$ & $-0.01 \pm 0.09$ \phn \\
Main component (gas) &  1.06 & 0.84 \phn\\
Sub component & $-0.05 \pm 0.10$ & $0.00\pm 0.07 $\phn \\
Sub component (gas) &  -0.73 & 0.10 \phn \enddata
\label{tab:offset}
\end{deluxetable}

\section{Possible systematic effects}
\label{sec:systematics}

To assess the reliability of the reconstructed map, we generate 1000
bootstrap resampled weak lensing catalogs and perform reconstructions
on each of these.  To further test the reliability of the strong
lensing data we create 10 different reconstructions each time
removing one of the multiply imaged systems we use. The resulting
$\kappa$-maps do not change substantially, the main features
(i.e. the ellipticity of the main and the subcluster) seen in
Fig.~\ref{fig:swunited} remain in all of the reconstructions.

We also perform a reconstruction using the weak lensing data only
(bottom-left panel of Fig.~\ref{fig:weakvsstrong}). From only weak
lensing, we clearly detect both the main and sub clusters, but are
unable to resolve individual substructures within each with high
significance. This failure can in part be attributed to the limited
field size used here in contrast to \citet{clowe06b} (our method at
present does not allow us to use larger fields for the reconstruction,
adaptive grid methods need to be employed then). In addition, the mass
estimates from weak lensing reconstruction are unreliable (due to the
the mass-sheet degeneracy); the resulting potential does not reproduce
any of the multiply imaged systems observed.  Traditionally, to
overcome the problem of the mass-sheet degeneracy and measure the mass
from the weak lensing mass reconstruction, one assumes a vanishing
surface mass density at the edge of the field or adopts parametrized
model fitting. The former is not applicable here because the field
used is too small.  The latter will yield reliable mass estimates if
the assumed profile is correct, but it is difficult to guess the
appropriate model for such a morphologically disturbed
cluster. Combining strong and weak lensing helps us to break this
degeneracy. However, the lack of counter arc candidates and redshifts
of multiply imaged systems limits our ability to provide even stronger
constraints on the mass distribution.

The errors on mass estimates quoted throughout the paper are obtained
by summing in quadrature the errors obtained from the reconstructions
using bootstrapping of the weak lensing catalogs and removing
individual strong lens systems.  In addition, we estimate and add
systematic uncertainties resulting from differences in the
reconstruction when using different initial models (as explained
below).  In principle, we should incorporate all of these effects
simultaneously in the reconstructions, however this is too
computational intensive and beyond the scope of this analysis.

\subsection{Initial conditions}
\label{sec:inicond}
Methods involving strong lensing mass reconstruction with many
multiple imaged systems are subject to degeneracies in parameter
space. It is difficult to search the parameter space because the
$\chi^2$ function has many local extrema. In our case, there are
pronounced, narrow secondary minima where individual systems are
reconstructed well by the model (and the others less so) and high
``walls'' in $\chi^2$ space where images happen to form close to the
corresponding critical curves. Because weak lensing data are noisy,
additional, though less pronounced, minima in $\chi^2$ are
created. Given that our reconstruction method does not guarantee that
we find a global minimum, in principle we would need to check the
whole parameter space. This search is unfeasible with any available
algorithms, given that the number of parameters (i.e., pixels in the
mass map) is $>1000$.

Instead, we opt to investigate the use of many initial models. We try
both extremes; i.e. models that are clearly in disagreement with the
data (zero initial conditions, single mass peak centered on the
position of the gas) and models that use alternative methods to
reconstruct the data (two-NIE model, two-NIE plus cluster members
model from strong lensing modeling, weak lensing mass model from
\citealp{clowe06b}). Although the details of the reconstruction depend
on the initial conditions (as can be seen from comparing all but
bottom-right $\kappa$-maps in Fig.~\ref{fig:weakvsstrong}), the main
features (SE-NW elongation of the main cluster, two mass peaks close
to the main cluster cDs, an elongation in the mass map toward the gas
peak, E-W elongation of the subcluster) are independent of the
particular initial model we use.

The resulting $\chi^2$ contribution of the weak lensing data
$\chi_{\epsilon}^2$ remains the same for all sets of initial
conditions ($\chi_{\epsilon}^2 \simeq 7000$ for $2N_{\rm gal} = 3800$
data points). The strong lensing contribution is lowest when
the two component NIE model is used ($\chi_{\rm M}^2 = 20$) and highest
when the zero model initial conditions ($\chi_{\rm M}^2 =
1000$) are used. Although the absolute values of the $\chi^2$ have little
meaning (i.e. one can change the regularization parameter to obtain a
lower value for $\chi_{\epsilon}^2$ because the weak lensing data is
noisy), the relative values tell us which resulting reconstruction is
indeed a better fit to the data (as the same regularization parameters
are used in all cases and it is the fit to the high signal to noise strong
lensing data that improves).

We estimate the systematic uncertainty on our mass estimates by
calculating the maximum difference in the resulting mass estimates
when using different initial models.  We exclude the zero model in
this calculation as it does not predict the strong lensing data
reliably. The resulting $\kappa$-map (see bottom-left panel in Fig.~\ref{fig:weakvsstrong}) 
for this model is suppressed
where only weak lensing data are available (due to the mass-sheet
degeneracy) and produce steep gradients close to arc A and a $\sim
10\%$ lower overall mass. However, other features in the
reconstruction (positions and shapes of the mass peaks for the main
and the sub-cluster) remain unchanged. In all cases, the resulting
mass peaks ``move'' away from the initial guess; showing that the
regularization is sufficiently flexible to reproduce the features in
the data. Therefore, we are confident that our results are not biased
as a result of the particular choices of regularization and are
independent of the initial conditions to the extent presented in
Fig.~\ref{fig:weakvsstrong}.

\subsection{Strong-lens system identification and redshifts}
\label{sec:iniconds}

Finally, the mass measurements presented above depend on the
reliability of the redshifts and identification used for strong
lensing reconstruction. These (except for system A) are not
secure and depend upon the model used in Sect.~\ref{sec:datasl}.  In
particular, the redshifts of the systems around the subcomponents (and
consequently the measurements of its mass) can change because the
subcomponent mass reconstruction does not depend strongly on the
modeling around the main component (where the arc with the known
redshift is located).

The parametrized model used in Sect.~\ref{sec:datasl} to estimate
the redshifts of the strong lensing systems likely deviates from the
truth model. We tested the dependence of the assumed redshifts by
performing 100 reconstructions in which we modified all of the strong
lensing system redshifts except for system A. For each reconstruction
these uncertain redshifts were changed by assuming a Gaussian scatter
with $\sigma_{\mathrm{z}}=0.5$ and requiring that all sources lie
behind the cluster. The resulting mass and position errors were
smaller (by a factor of $\sim$2) than the ones quoted throughout the
paper, and the mass and position measurements were unbiased (as
expected from such an analysis). The mass measurements would change,
however, if we systematically increase (decrease) the redshifts of all
the systems. For example, if we decrease the redshifts of the
subcluster systems such that the lowest redshift system of the
subcluster (G) is at $z_{\rm s}=0.5$ (and scale the others
accordingly), then the mass increases by a factor of $\sim
Z(1.3)/Z(0.5)=2$. If we instead increase all the redshifts, pushing
the highest redshift system (I) to $z=\infty$), then the mass of the
subcluster will decrease by $1.0/Z(2.1) = 1.2$. Note however that
these are just estimates, since the redshift estimates of the weak
lensing sources also plays a role, and the changes will therefore be
even smaller.

As a further test, we also perform 10 reconstructions where we remove
one strong lens system at a time from the analysis. In
Fig.~\ref{fig:slsyst} we show the two that differ the most from the
reconstruction that uses all the strong lensing systems. When we
remove system I from the reconstruction (right panel in
Fig.~\ref{fig:slsyst}) the double peaked structure of the subcluster
disappears, while the east-west elongation remains. For the main
cluster we see the largest deviations from the original map when we
remove system D from the reconstruction (left panel in
Fig.~\ref{fig:slsyst}). However, these results do not mean that these
structures are not real, rather they illustrate some possible
systematics when dealing with sparse strong lensing data.  We are
confident that system D is indeed multiply imaged, and images I2-I3,
at least, are also probably multiply imaged.

These tests demonstrate that even though the strong lensing system
identification and their redshift determinations are important to the
details of the reconstruction, the errors arising from
misidentifications/redshift misestimation are within those quoted in
the previous section.

\subsection{Predictive power of the model}
\label{sec:predict}

The predicted redshifts and strong lensing system identification can
be tested using spectroscopy. We are planning just
such a survey and will therefore be able to test the fidelity of our
model in the future.  In addition, using Spitzer IRAC data we identify a
candidate high redshift object.  While bright in all IRAC bands, the
faintness of this galaxy in optical and near infrared bands indicates
that it could lie at $z \gtrsim 6$. It has two components and our
predicted critical curve for this redshift (indistinguishable from the
$z\to\infty$ critical curve) is exactly where it should be if indeed
these sources are merging images of the same source; they are denoted
with crosses and plotted together with the critical curve in
Fig.~\ref{fig:ccurve}. Furthermore, we  identify  a candidate
counter image with similar colors in the IRAC data just outside the
critical curve. All three need to be further analyzed and
confirmed (using spectroscopy and/or NIR data) before we include
them in the modeling.

We also identify possible counter-images for systems B, J, H, and I as
predicted from our model, they are indicated in
Fig.~\ref{fig:predict}. If these and high-redshift candidate are
confirmed, the predictive power of our model would further indicate
that the strong lensing data we use is reliable enough for this work.

\begin{figure*}[ht]
\begin{center}
\includegraphics[width=1.0\textwidth]{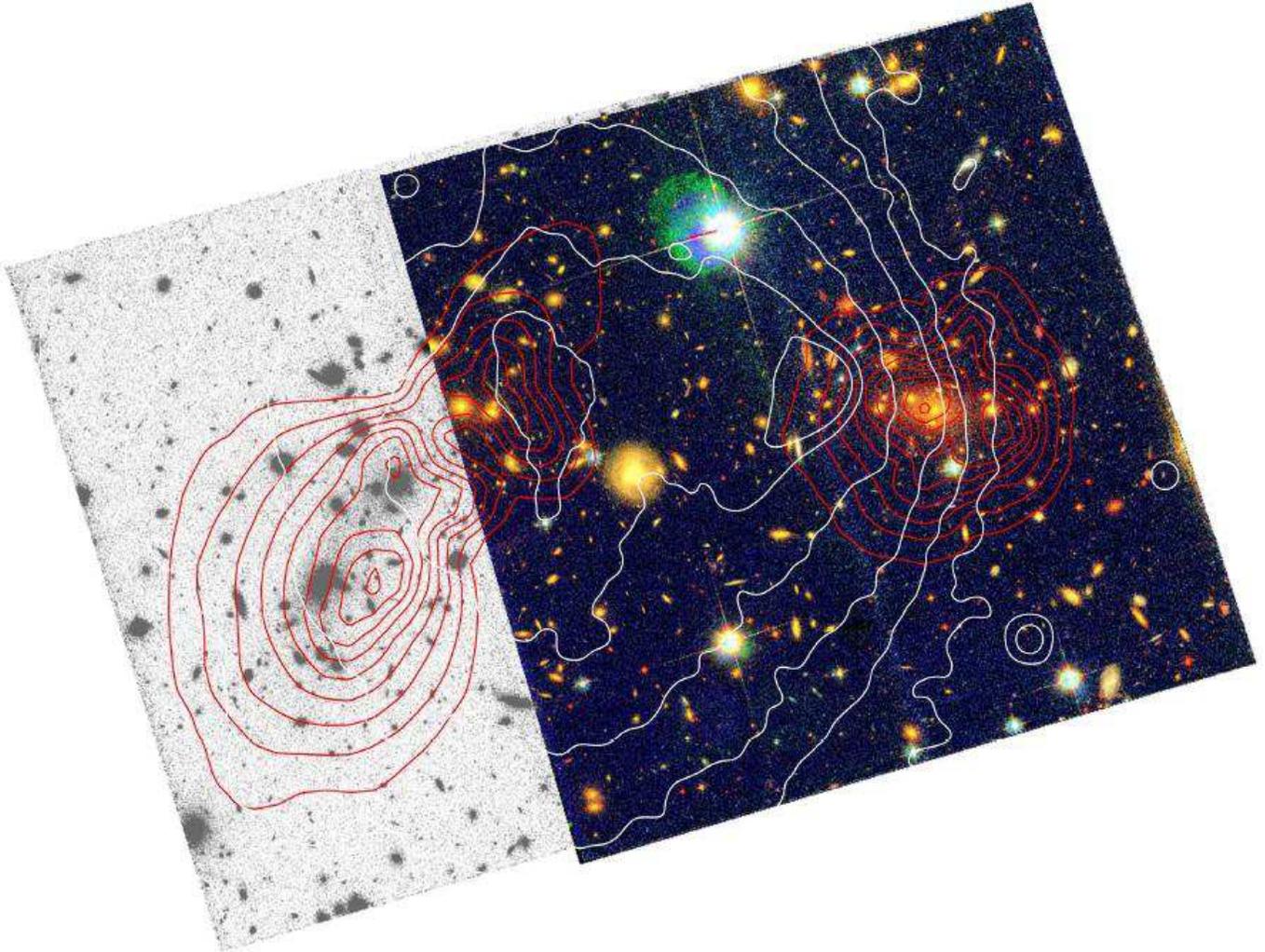}
\end{center}
\caption{The F435W-F606W-F814 color composite of the cluster
  \protect \bullet. Overlaid in {\it red contours} is the surface mass
  density $\kappa$ from the combined weak and strong lensing mass
  reconstruction (for the purpose of this plot we recalculate the
  final $\kappa$-map from top-left panel of Fig.~\ref{fig:weakvsstrong} on a finer
  grid). The contour levels are linearly spaced with $\Delta\kappa =
  0.1$, starting at $\kappa=0.5$, for a fiducial source at a redshift
  of $z_{\rm s} \to \infty$. The X-ray brightness contours from the
  500 ks Chandra ACIS-I observations \citep{markevitch06} are overlaid
  in {\it white}. North is up and East is left, the field is
  $4.9^{\prime}\times 3.2^{\prime}$, which corresponds to $1300 \times
  830 \mbox{ kpc}^2$ at the redshift of the cluster. The color
  composite was created following the algorithm from
  \citet{lupton04}.}
\label{fig:swunited}
\end{figure*} 
\begin{figure}[ht]
\begin{center}
\includegraphics[width=0.5\textwidth]{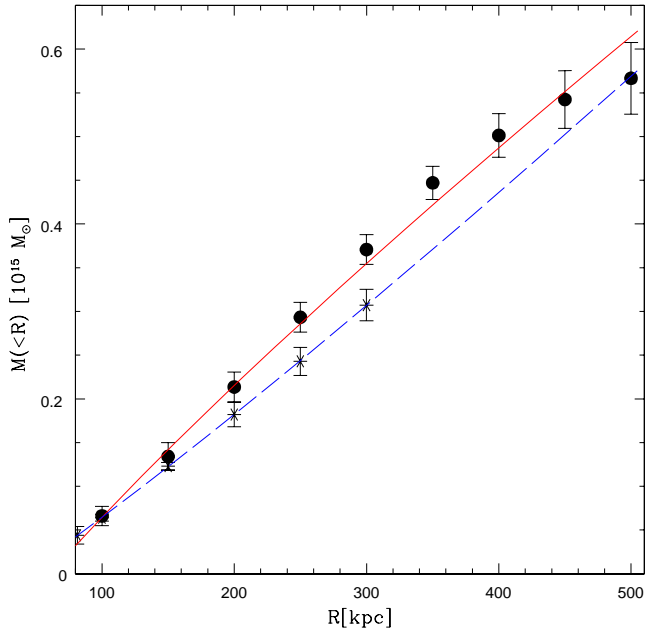}
\end{center}
\caption{The integrated mass profile of \protect \bullet. The profile was
determined by measuring the enclosed mass in cylinders, centered on the
southern cD of the main cluster (dots), and the BCG of the subcluster
(crosses). We fit power law profiles to the data $M(<R) \propto
R^{n_{\rm i}}$, the results are shown for the main (solid line) and
for the subcluster (dashed line).}
\label{fig:profile}
\end{figure} 

\begin{deluxetable*}{rrrrr} 
\tablecolumns{5} 
\tablewidth{0pc} 
\tablecaption{Comparison between the
gas mass and the total mass for four elliptical regions centered on
the maxima of the gas density map and the southern cD / BCG of the main / 
subcluster. $R_a$ and $R_b$ indicate semi-major 
and semi-minor axes ($R_a$ is oriented w.r.t. RA) of the region considered. Last row gives the
corresponding numbers for the entire ACS field (rectangle).}  
\tablehead{
\colhead{} & \colhead{$R_a \times R_b [\mbox{kpc}] ([\mbox{arcmin}])$}
& \colhead{$M_{\rm gas} [10^{13} M_{\odot}]$} & \colhead{$M_{\rm
total} [10^{13} M_{\odot}]$} & \colhead{ $M_{\rm gas} / M_{\rm
total}$}} 
\startdata Main (gas peak) & & 2.0 $\pm$ 0.2 &
10.8$\pm$ 0.6 & 0.19$\pm$ 0.03\phn\phn\\ 
Main (S cD) & \raisebox{1ex}{125 (0.47) $\times$ 250 (0.59)} & 1.4 $\pm$ 0.1 & 16.3 $\pm$ 0.9 & 0.09 $\pm$ 0.01 \phn \\ 
\cline{1-5} 
Sub (gas peak) &
& 0.42 $\pm$ 0.04 & 2.1$\pm$ 0.2 & 0.20 $\pm$ 0.06\phn\phn \\ 
Sub (BCG) & \raisebox{1ex}{80 (0.31) $\times$ 80 (0.31)} & 0.18 $\pm$
0.02 & 4.3 $\pm$ 0.2 & 0.04 $\pm$ 0.01 \phn \\ 
\cline{1-5} 
ACS field& 1300 (4.9) $\times$ 830 (3.2) & 13 $\pm$ 1 & 91 $\pm$ 10 & 0.14 $\pm$ 0.03 \phn \enddata
\label{tab:massgas}
\end{deluxetable*}

\begin{figure*}[ht]
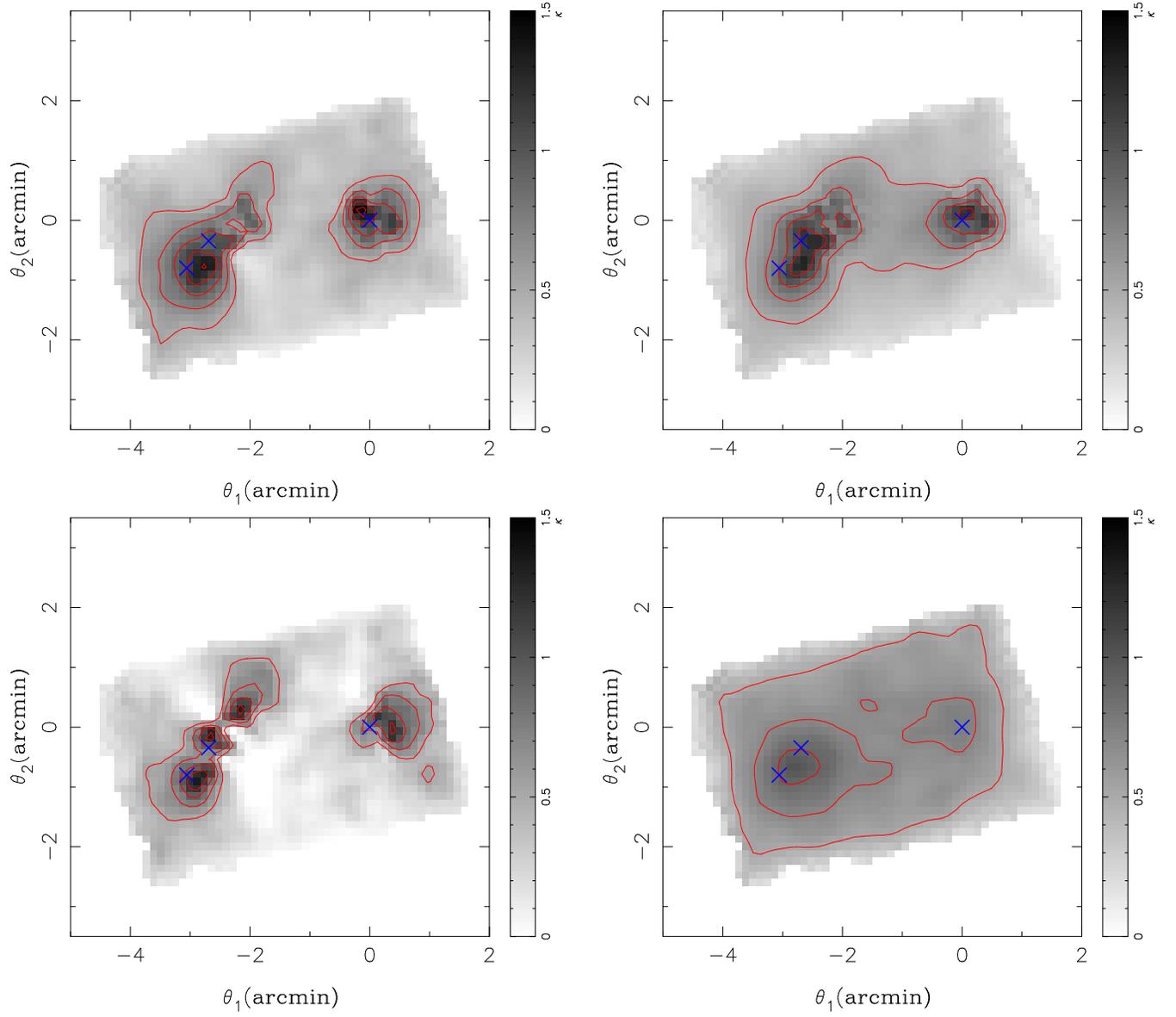

\begin{minipage}{0.5\textwidth}
\centerline{\includegraphics[width=0.95\textwidth]{f6a.eps}}
\end{minipage}
\begin{minipage}{0.5\textwidth}
\centerline{\includegraphics[width=0.95\textwidth]{f6b.eps}}
\end{minipage}
\begin{minipage}{0.5\textwidth}
\centerline{\includegraphics[width=0.95\textwidth]{f6c.eps}}
\end{minipage}
\begin{minipage}{0.5\textwidth}
\centerline{\includegraphics[width=0.95\textwidth]{f6d.eps}}
\end{minipage}
\caption{The effects of different initial conditions (a-c) and
  exclusion of the strong lensing data (d) on the final
  reconstruction. Shown is the surface mass density $\kappa$ of the
  weak and strong lensing reconstruction using the two component NIE
  model ({\it top-left}) as initial guess, using weak lensing reconstruction
  from \citet{clowe06b} ({\it top-right}), and using zero
  initial model ({\it bottom-left}).  The weak lensing only mass
  reconstruction is shown in ({\it bottom-right}); all for a fiducial source at
  a redshift of $z_{\rm s} \to \infty$. The contour levels are
  linearly spaced with $\Delta\kappa = 0.2$, starting at $\kappa=0.5$.
  The two cD's and the BCG of the subcluster are denoted with
  crosses.}
\label{fig:weakvsstrong}
\end{figure*}

\begin{figure*}[ht]
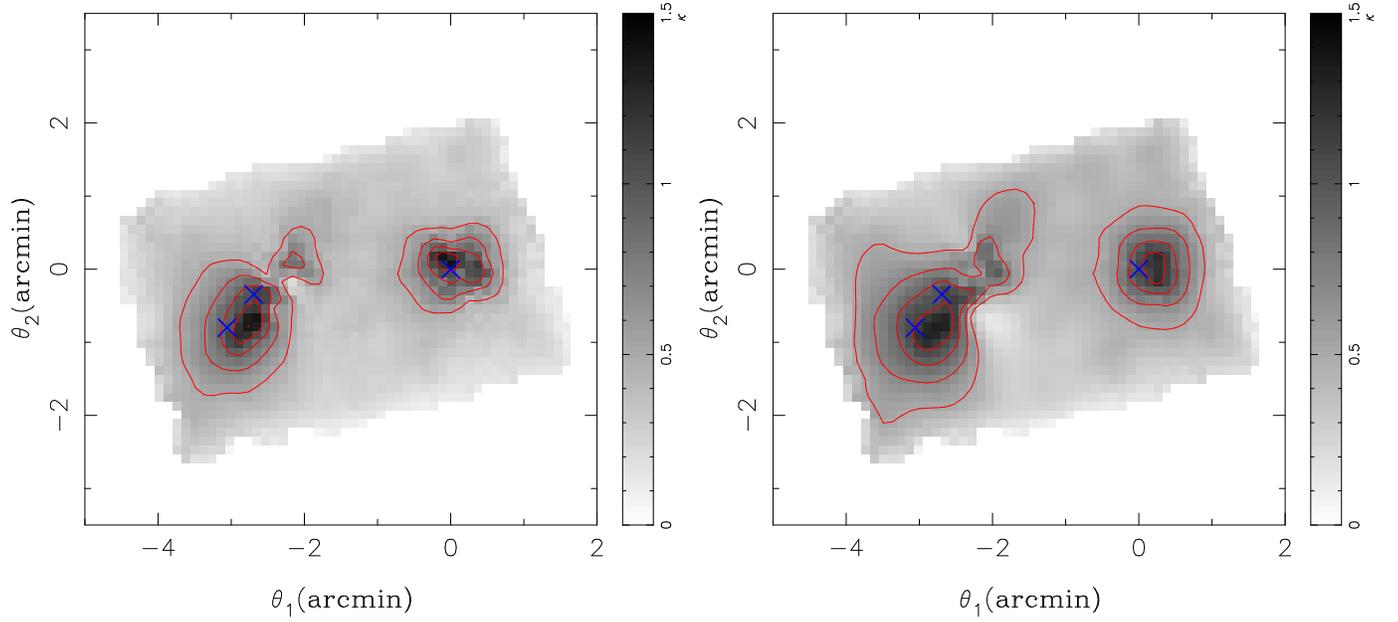

\begin{minipage}{0.5\textwidth}
\begin{center}
\includegraphics[width=1.0\textwidth]{f7a.eps}
\end{center}
\end{minipage}
\begin{minipage}{0.5\textwidth}
\begin{center}
\includegraphics[width=1.0\textwidth]{f7b.eps}
\end{center}
\end{minipage}
\caption{The surface mass density $\kappa$ of the two reconstructions
  where we  ({\it left}) removed image system D, and  ({\it right}) image system I from the data. These are two extreme examples where the
  resulting $\chi^2$ decreased at most; still the resulting maps are
  very similar to the original reconstruction. The contour levels are
  the same as in Fig.~\ref{fig:weakvsstrong}, the two cD's of the main and the BCG of the subcluster are denoted
  with crosses.}
\label{fig:slsyst}
\end{figure*} 

\begin{figure*}[ht]
\begin{center}
\includegraphics[width=0.8\textwidth]{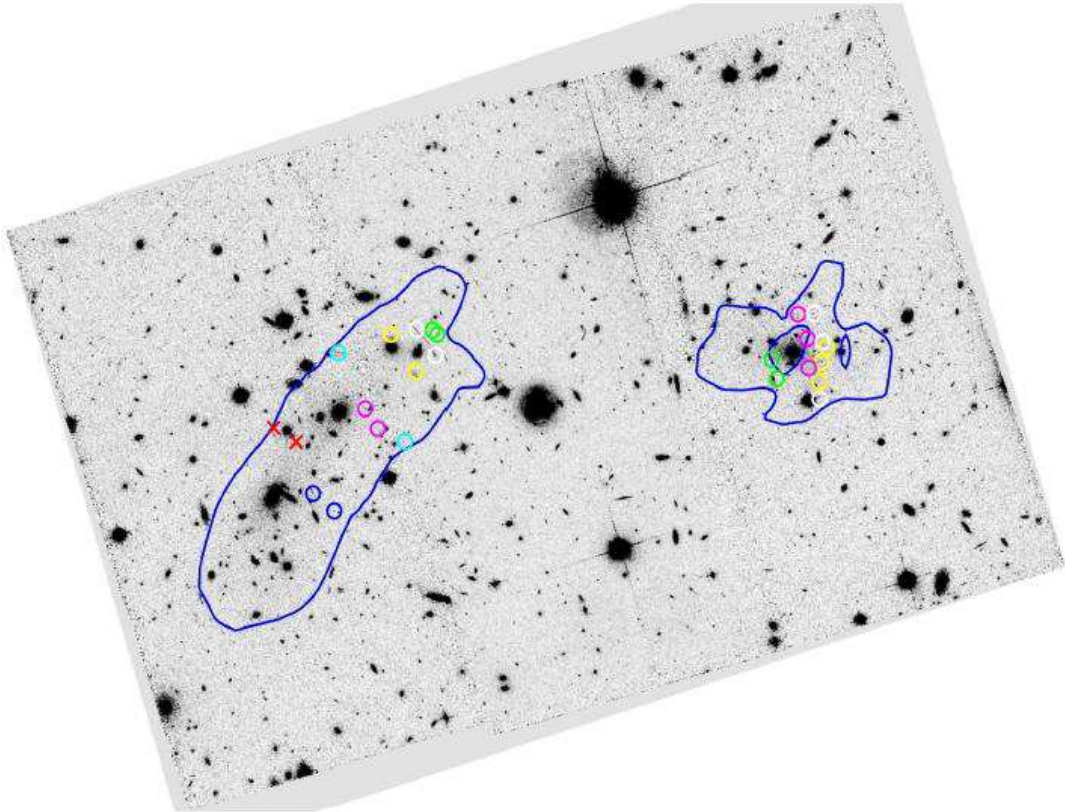}
\end{center}
\caption{The critical curve for a fiducial source at $z\to\infty$ from
the combined strong and weak lensing reconstruction. Circles denote
the positions of the multiple image systems we used, while the crosses
denote the candidate high redshift source identified using Spitzer
data (note that this system was {\it not} included in the
reconstruction). North is up and East is left, the field is
$4.9^{\prime}\times 3.2^{\prime}$.}
\label{fig:ccurve}
\end{figure*}

\begin{figure*}[ht]
\begin{minipage}{0.5\textwidth}
\begin{center}
\includegraphics[width=1.0\textwidth]{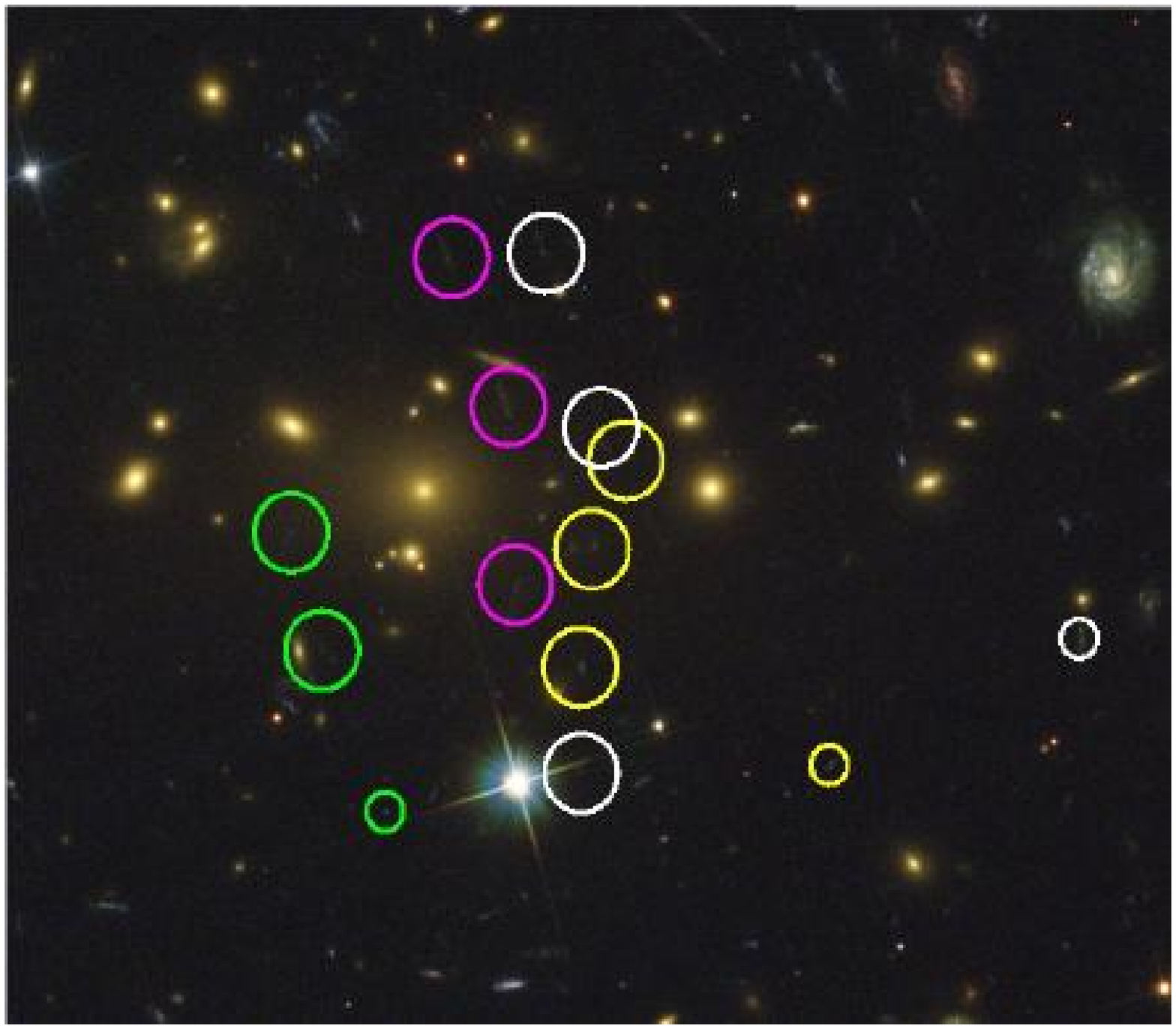}
\end{center}
\end{minipage}
\begin{minipage}{0.5\textwidth}
\begin{center}
\includegraphics[width=1.0\textwidth]{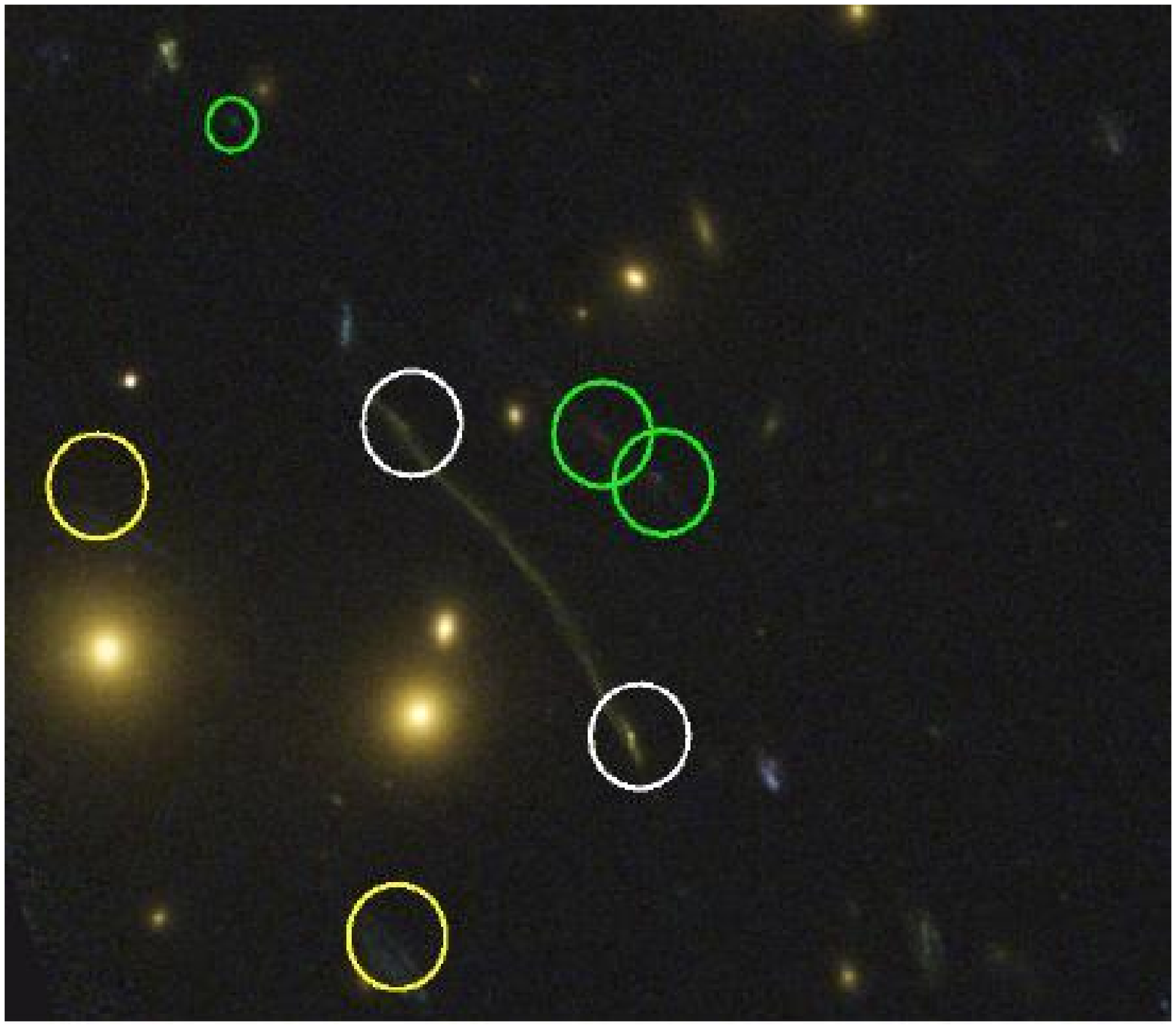}
\end{center}
\end{minipage}
\caption{The F435W-F606W-F814W color composite of the $1.1^{\prime}\times 0.9^{\prime}$ inner part of the
subcluster ({\it left}) and of the $0.6^{\prime} \times 0.5^{\prime}$ cutout around arc B ({\it right}) with large
circles denoting the positions of the multiple image systems we
used. The small circles show the candidate counter images as
predicted by the weak and strong lensing reconstruction (not used in
the final reconstruction). Image courtesy of Mischa Schirmer.}
\label{fig:predict}
\end{figure*} 


\section{Conclusions and future work}
\label{sec:conclusions}

Massive and interacting clusters, while quite rare, are
remarkably well-suited to addressing outstanding issues in both galaxy
evolution and fundamental physics. However, in order to study the mass
distribution, methods relying on hydrostatic (X-rays)
or dynamical equilibrium are ill-suited for such systems.

We have applied a mass reconstruction method based on strong and weak
gravitational lensing to the cluster {\bullet}. We use deep, high
resolution optical data to identify objects belonging to the same
multiple-image systems.  The same data are used to obtain weak lensing
catalogs allowing us to obtain a strong+weak lensing mass map of the
cluster core.  Our main conclusions are the following:
\begin{enumerate}
\item 
Using the combined strong and weak lensing mass reconstruction we
derive a high-resolution, absolutely calibrated mass map; we get
projected, enclosed mass $M_{\rm main}(<250\:\mbox{kpc})= (2.8 \pm
0.2) \times 10^{14} M_{\odot}$ around the main and $M_{\rm
sub}(<250\:\mbox{kpc})= (2.3 \pm 0.2) \times 10^{14} M_{\odot}$ around
the subcluster.
\item We detect the main cluster peak and a distinct mass
concentration at the subcluster position, both clearly offset from the
location of the X-ray gas in the system (at 10-$\sigma$ and
6-$\sigma$ significance for the main and the subcluster respectively - see
Fig.~\ref{fig:swunited} and Tab.~\ref{tab:offset}).
\item The majority of the mass is spatially coincident with the
galaxies, which implies that the cluster mass must be dominated by a
relatively collisionless form of dark matter. The high resolution data
allow us to significantly detect the shapes of both
the main mass component and the subcluster with no prior assumptions
on their positions or profiles.
\end{enumerate}
We show that the cluster {\bullet} is a very efficient strong lens;
the area enclosed within the critical curve for highest redshift
sources is comparable to that of A1689 \cite{broadhurst05}.
Consequently, we plan to conduct a pencil beam survey for
high-redshift ($z\sim 7$) objects. At present this is not possible,
because we lack both reliable redshift information for the {\bullet}
multiple image systems, and multi-color high-resolution imaging data
in the main cluster region. The latter is needed to identify counter
images for the strong lens systems, and so improve the reliability of
the mass map to the south-east of the main cluster.  Once these data
are obtained (HST Cycle 15), we will improve our methodology to
include an adaptive grid approach, which will greatly improve the
constraints on the location of critical curves for these high-redshift
sources.

The prospects for such a search are however very promising, as we
already have two strong candidate high redshift objects. One is the
red arc B, which is constrained to lie at $3.24 < z < 6$ based upon
the redshift of arc A and detection in F814W. The other (see
Fig.~\ref{fig:ccurve}) is Spitzer IRAC-selected with a likely redshift
of $z \gtrsim 6$. Using this cluster as a gravitational telescope will
therefore allow us to study galaxy properties near the epoch of
reionization.

The high resolution mass map of this cluster, in addition to enabling
the high-z galaxy search, will be valuable for several other ongoing
science programs.  Most significantly, we are currently running
simulations to reconstruct the dynamical history of the cluster
merger, and provide an independent constraint on the dark matter
self-interaction cross-section. The relative masses of the main and
the subcluster, and the central mass distribution, two critical
parameters for the simulations, were the limiting source of
uncertainty prior to this work.


\begin{acknowledgements}
 We would like to thank Marco Lombardi, Roger Blandford, Peter
 Schneider, and Steve Allen for many useful discussions that helped to
 improve the paper, and Mischa Schirmer for producing the color image
 used for Fig.~\ref{fig:predict}. We are grateful to the anonymous
 referee for valuable comments and suggestions that helped to improve
 this manuscript, and for pointing out a typo in our previous paper of
 the series. Further we would like to thank Volker Springel for
 providing us with the simulations used in the paper. Support for this
 work was provided by NASA through grant number HST-GO-10200.05-A from
 the Space Telescope Science Institute, which is operated by AURA,
 Inc., under NASA contract NAS 5-26555. MB acknowledges support from
 the NSF grant AST-0206286. This project was partially supported by
 the Department of Energy contract DE-AC3-76SF00515 to SLAC. DC and DZ
 acknowledge support from NASA through grant number
 LTSA04-0000-0041. MM acknowledges support from Chandra grant
 GO4-5152X.
\end{acknowledgements}


\end{document}